\documentclass[usenatbib]{mnras}
\usepackage{graphicx, amssymb, natbib, bm, aas_macros, pdflscape}
\usepackage{ marvosym }
\usepackage{enumitem}
\usepackage{xcolor}
\usepackage[normalem]{ulem}

\newcommand{\rvir}{R_{\rm{vir}}}

\newcommand{\mstar}{{M}_{\star}}

\newcommand{\msun}{{\rm M}_{\odot}}
\newcommand{\lsun}{L_{\odot}}

\newcommand{\bi}{\begin{itemize}}
\newcommand{\ei}{\end{itemize}}

\definecolor{myorange}{RGB}{224, 97, 33}

\newcommand*{\img}[1]{%
    \raisebox{-.1\baselineskip}{%
        \includegraphics[
        height=\baselineskip,
        width=\baselineskip,
        keepaspectratio,
        ]{#1}%
    }%
}

\title[Quenching of Ultra-Faint Dwarfs with Fat ELVIS \img{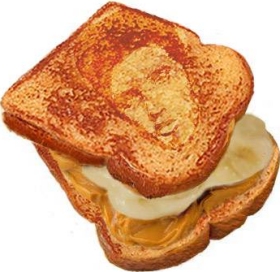}]
{The Suppression of Star Formation on the Smallest Scales: What Role Does Environment Play?}

\author[Rodriguez Wimberly et al.]
{M.~K.~Rodriguez Wimberly,$^1$\thanks{$\!\!$e-mail: wimberlm@uci.edu}
M.~C.~Cooper,$^1$\thanks{$\!\!$e-mail: cooper@uci.edu} 
S.~P.~Fillingham,$^1$
\newauthor 
M.~Boylan-Kolchin,$^2$
J.~S.~Bullock,$^1$
S.~Garrison-Kimmel$^3$\thanks{$\!\!$Einstein fellow} 
\\
$\!\!^1$Center for Cosmology, Department of Physics \& Astronomy,
4129 Reines Hall, University of California, Irvine, CA 92697, USA \\
$\!\!^2$Department of Astronomy, The University of Texas at Austin,
2515 Speedway, Stop C1400, Austin, TX 78712, USA \\
$\!\!^3$TAPIR, Mailcode 350-17, California Institute of Technology,
Pasadena, CA 91125, USA}

\begin{document}

\pagerange{\pageref{firstpage}--\pageref{lastpage}} 
\pubyear{2017}

\maketitle

\label{firstpage}
\begin{abstract}
  
  The predominantly ancient stellar populations observed in the lowest-mass
  galaxies (i.e.~ultra-faint dwarfs) suggest that their star formation was
  suppressed by reionization. Most of the well-studied ultra-faint dwarfs, however, are
  within the central half of the Milky Way dark matter halo, such that they are
  consistent with a population that was accreted at early times and thus
  potentially quenched via environmental processes. To study the potential role
  of environment in suppressing star formation on the smallest scales, we
  utilize the Exploring the Local Volume in Simulations (ELVIS) suite of
  $N$-body simulations to constrain the distribution of infall times for
  low-mass subhalos likely to host the ultra-faint population. For the
  ultra-faint satellites of the Milky Way with star-formation histories inferred
  from {\it Hubble Space Telescope} imaging, we find that environment is highly
  unlikely to play a dominant role in quenching their star formation. Even when
  including the potential effects of pre-processing, there is a $\lesssim 0.1\%$
  probability that environmental processes quenched all of the known ultra-faint
  dwarfs early enough to explain their observed star-formation
  histories. Instead, we argue for a mass floor in the effectiveness of
  satellite quenching at roughly $\mstar \sim 10^{5}~\msun$, below which star
  formation in surviving galaxies is globally suppressed by
  reionization. We predict a large population of quenched ultra-faint
    dwarfs in the Local Field ($1 < R/\rvir < 2$), with as many as $\sim 250$ to
    be discovered by future wide-field imaging surveys.

\end{abstract}


\section{Introduction}
\label{sec:intro} 

The Local Group serves as a cosmic Rosetta Stone, offering the opportunity to
study galaxy formation and evolution at a level of detail not possible at
cosmological distances \citep{bk16}. This is especially true at the smallest
galactic scales --- i.e.~for very low-mass galaxies or what are often referred
to as ultra-faint dwarfs (UFDs). Photometric observations of UFDs in the Local
Group find universally old stellar populations, such that these systems have
typically ceased forming stars by $z \sim 2$ \citep[or a lookback time of
$\sim10.3$~Gyr,][]{brown14, weisz14a}. The prevalence of ancient stellar
components in these extremely low-mass systems is commonly interpreted as
evidence of star formation suppression via reionization, where a photoionizing
background increases the cooling time for low-density gas so as to quell the
fuel supply for star formation in the lowest-mass halos
\citep[e.g.][]{efstathiou92, quinn96, thoul96}.

\begin{figure*} 
\centering 
\includegraphics[width=0.9\textwidth]{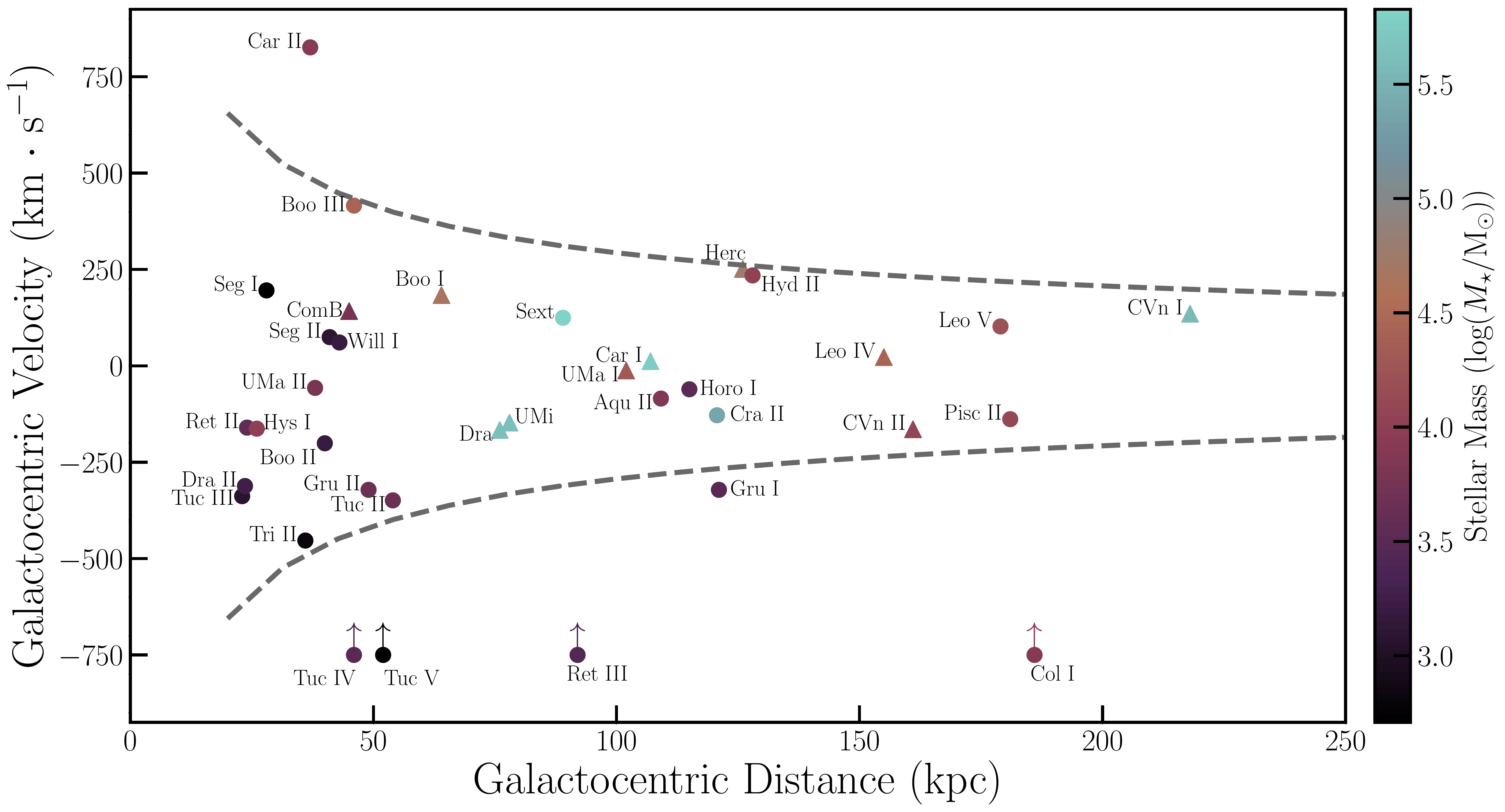}
\caption{Galactocentric velocity versus distance for the sample of UFD
  satellites of the Milky Way. Points are color-coded according to stellar mass,
  assuming a $V$-band mass-to-light ratio of 1.2; the triangles denote those
  objects with a published SFH from \citet{brown14} or \citet{weisz14a}. To
  account for unknown tangential motion, the observed line-of-sight velocities
  have been multiplied by a factor of $\sqrt{3}$. Those systems without
  published line-of-sight velocity measurements (Tuc~IV, Tuc~V, Ret~III, and
  Col~I) are plotted at
  $\sqrt{3} \cdot V_{\rm \ell os} = -750~{\rm km}~{\rm s}^{-1}$ with upward
  arrows representing the uncertainty in their $V_{\ell \rm os}$. Masses
  (i.e.~luminosities), distances, and line-of-sight velocities for this sample
  are based on published values from \citet{mcconn12}, \citet{des1},
  \citet{des2}, \citet{laevens15}, \citet{brown14}, \citet{weisz14a},
  \citet{simon15, simon17}, \citet{kirby15a}, \citet{kirby13, kirby15b,
    kirby17}, \citet{li18}, \citet{torrealba16}, \citet{torrealba18},
  \citet{caldwell17}, \citet{martin16b, laevens15b}, \citet{walker16},
  \citet{koposov18}, and references therein.}
\label{fig:ufdsamp} 
\end{figure*}

While the measured star formation histories of UFDs are broadly compatible with
quenching via reionization, the most well-studied systems in the Local Group are
located at relatively small galactocentric radii, which is also consistent with
a population that was accreted at early cosmic time \citep{rocha12, oman13}.
As such, the old stellar populations identified in UFDs orbiting the Milky Way
and M31 may instead be the result of environmental processes that quenched star
formation following infall onto the host halo.
For example, recent measurements of the proper motion for the Segue~I dwarf
\citep{belokurov07} suggest that it was accreted by the Milky Way halo roughly
$9.4$~Gyr ago \citep{fritz17}, such that rapid environmental quenching would
produce an ancient and metal-poor stellar population as observed today
\citep{frebel14, webster16}.
Undoubtedly, observations of isolated UFDs (i.e.~beyond the reach of
environmental effects) would be an excellent way to differentiate between these
two physical scenarios (quenching via reionization versus via environment).
Current datasets, however, lack the depth to identify and characterize the
stellar populations of UFDs in the local field.

To address the potential role of environment in quenching UFDs, we utilize a suite of $N$-body simulations to track the
accretion and orbital history of the low-mass subhalos that host the UFD
satellite population.
Our approach is similar to that utilized by \citet[][see also
\citealt{weisz15}]{rocha12}, with the clear distinction that we aim to study the
UFD satellites as an ensemble and not on an object-by-object basis. For example,
herein, we study the likelihood that the $6$ galaxies in the UFD sample from
\citet{brown14} were accreted at early cosmic times, such that environmental
quenching could reproduce their observed star formation histories.
Overall, we strive to quantify the likelihood that environmental effects can
explain the universal ancient stellar populations in the lowest-mass galaxies.
In \S\ref{sec:dat}, we provide a brief census of the UFD satellite population of
the Milky Way along with a description of our simulation dataset and our primary
analysis methods. In \S\ref{sec:results}, we present our results regarding the
role of environment in quenching UFDs. Finally, we conclude with a brief
discussion and summary of our work in \S\ref{sec:dis} and \S\ref{sec:sum},
respectively.

\section{Data}
\label{sec:dat}

\subsection{UFD Galaxy Sample}
\label{sec:ufd}

Since the discovery of the first ultra-faint dwarfs using photometric data from
the Sloan Digital Sky Survey \citep[SDSS,][]{york00}, a large number of UFDs
have been identified as satellites of the Milky Way \citep[e.g.][]{willman05a,
  willman05b, zucker06a, zucker06b, belokurov10, des1, des2}.
Deep imaging of M31 has likewise uncovered a population of UFDs
orbiting M31, with similarly old stellar populations \citep[e.g.][]{martin09,
  weisz14a, skillman17}.
Throughout this work, we focus our analysis on the ultra-faint satellite
population of the Milky Way, selecting all systems with
$L_{V} < 5 \times 10^{5}~\lsun$ ($M_{V} > -9.3$) as UFDs.
Figure~\ref{fig:ufdsamp} shows the position and line-of-sight velocity of these
systems relative to the Milky Way, with velocities scaled by a factor of
$\sqrt{3}$ to crudely account for potential tangential motion.\footnote{This
  typically serves as a lower limit to the total velocity, with the recently
  measured motions for a subset of UFDs from Gaia Data Release 2 \citep{simon18,
    fritz18} yielding higher total velocities than our $\sqrt{3}V_{\ell os}$
  estimate.}

Of the $36$ known UFD satellites of the Milky Way, there are published
star-formation histories (SFHs) in the literature for $10$ based on {\it Hubble
  Space Telescope} ({\it HST}) imaging from \citet{brown14} and
\citet{weisz14a}.
For all $10$ of these systems, the reported mean stellar age is $>9~{\rm Gyr}$
with $90\%$ of the stars forming by $z \sim 2$. For the small number of objects
included in both the \citet{brown14} and \citet{weisz14a} samples, there is
relatively good agreement between the measured SFHs.
The exception is CVn~II, for which \citet{weisz14a} find a tail of star
formation extending to $z \sim 1$. The {\it HST}/WFPC2 imaging analyzed by
\citet{weisz14a}, however, is shallower and covers a smaller area than the {\it
  HST}/ACS imaging utilized by \citet{brown14}, such that greater photometric
errors may be increasing the dispersion in the main sequence turn-off population
and thereby yielding a broader SFH.
Altogether, observations of the known UFD population orbiting the Milky Way
suggest that these very low-mass systems have old stellar populations, with
little star-formation activity since $z \sim 1-2$ \citep[e.g.][]{okamoto08,
  okamoto12, dejong08, sand09, sand10, sand12, brown12, martin15, bettinelli18}.

\begin{figure*}
\centering
\includegraphics[width=0.95\textwidth]{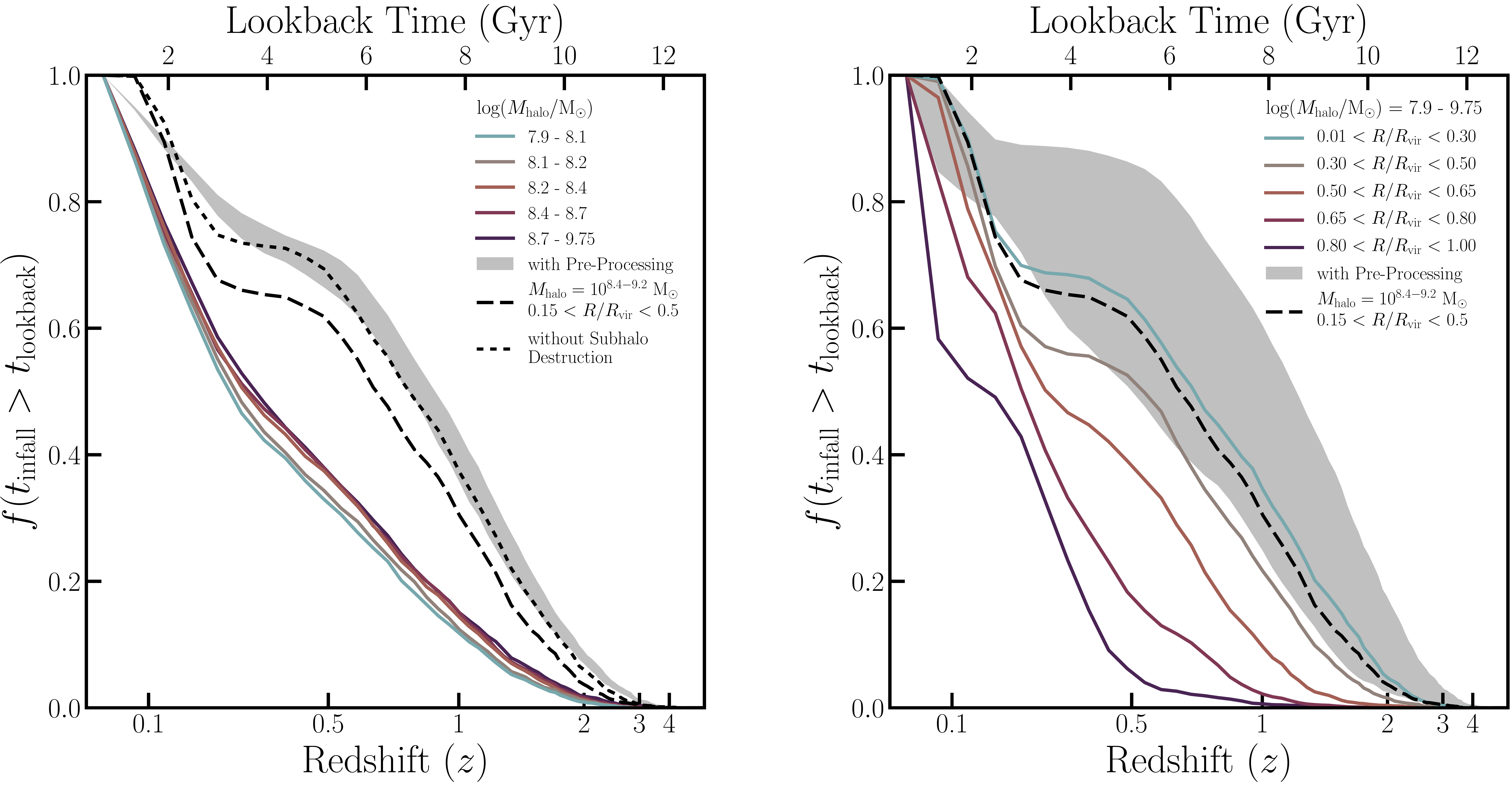}
\caption{The cumulative distribution of infall times ($t_{\rm infall}$) as a
  function of redshift for subhalos likely to host the Milky Way UFD satellite
  population. In both panels, the black long-dashed line corresponds to the
  distribution of infall times for our fiducial selection criteria, where
  subhalos are restricted to $0.15 < R < 0.50~R_{\rm vir}$ and
  $M_{\rm peak} = 10^{8.4 - 9.2}~\msun$. At \emph{left}, we show the variation
  in infall times as a function of subhalo mass, for all subhalos within
  $R_{\rm vir}$ at $z = 0$. At \emph{right}, we plot the infall time
  distribution across bins in host-centric distance for all subhalos with
  $M_{\rm peak} = 10^{7.9-9.75}~\msun$.  Each bin in distance or mass contains
  an approximately equal number of subhalos ($N \sim 3050$). The dotted line
  illustrates the distribution of infall times for our fiducial sample without
  including the effects of subhalo disruption (i.e.~using the original ELVIS
  catalogs versus the Fat ELVIS catalogs). Finally, the grey bands illustrate
  the corresponding distributions for infall onto a $\geq$ SMC-like host halo
  prior to the last infall onto a Milky Way-like host halo (see
  \S\ref{sec:sim}). While the distribution of infall times is largely
  independent of subhalo mass (and thus our assumed stellar mass-halo mass
  relation), it is strongly dependent upon host-centric (i.e.~galactocentric)
  distance.}
\label{fig:cumlaccrt}
\end{figure*}

\subsection{\emph{N}-Body Cosmological Simulations}
\label{sec:sim}

To investigate the role environmental mechanisms play in the quenching of UFDs,
we utilize the Exploring the Local Volume In Simulations (ELVIS) suite of $36$
high-resolution, cosmological zoom-in simulations of Milky Way-like halos
\citep{gk14}. Within the suite, $24$ simulations are of isolated Milky Way-like
halos and 12 are of Milky Way- and M31-like pairs. Each simulation occurs within
a high-resolution uncontaminated volume spanning $2-5$~Mpc in size with a
particle mass of $1.9 \times 10^{5}~\msun$ and a Plummer-equivalent force
softening length of $\epsilon = 141$~physical parsecs. Within the
high-resolution volumes, the halo catalogs are complete down to
$M_{\rm halo}>2 \times 10^{7}~\msun$, $V_{\rm max}>8~{\rm km}~{\rm s}^{-1}$,
$M_{\rm peak}>6 \times 10^{7}~\msun$, and
$V_{\rm peak}>12~{\rm km}~{\rm s}^{-1}$ --- thus sufficient to track the
evolution of halos hosting Local Group dwarfs with stellar masses of
$\sim10^{3-5}~\msun$. ELVIS adopts a $\Lambda$CDM cosmological model based on
\emph{Wilkinson Microwave Anisotropy Probe} 7-year data \citep{komatsu11,
  larson11} with the following parameters: $\sigma_{8} = 0.801$,
$\Omega_{\rm m} = 0.266$, $\Omega_{\Lambda} = 0.734$, $n_{\rm s} = 0.963$, and
$h = 0.71$.

As a dark matter-only simulation suite, ELVIS fails to capture the impact of the
host baryonic component on the subhalo population. In short, the inclusion of a
disk potential can substantially alter the subhalo distribution inside of the
host virial radius by tidally disrupting subhalos \citep{donghia10, brooks13, brooks14,
  gk17, sawala17}.
This subhalo destruction preferentially occurs in objects with early infall
times and/or more radial orbits.
As such, the distribution of subhalo infall times for a dark matter-only
simulation (such as ELVIS) will be biased towards earlier cosmic times, so as to
overestimate the role of environmental mechanisms in quenching star formation at
high $z$.

To account for the impact of the host baryonic component, following the work of
\citet{fham18}, we implement a correction to the ELVIS subhalo population that
will broadly capture the tidal effects of the host.
Based on Figures~5 and A2 from \citet{gk17}, we model the ratio of subhalos in
dark matter-only versus hydrodynamic simulations of Milky Way-like hosts as
$$  N_{\rm DMO} / N_{\rm HYDRO} =  40 \, e^{-22 \, d_{\rm peri}/{\rm kpc}} \ ({\rm
  for}~d_{\rm peri} < 50~{\rm kpc}), $$
where $N_ {\rm DMO}$ is the number of subhalos that survive to present-day in
a dark matter-only simulation, $N_ {\rm HYDRO}$ is the corresponding subhalo
count for a hydrodynamic simulation, and $d_{\rm peri}$ is the host-centric
distance at pericenter in kpc.
This relationship between pericentric passage and the likelihood of subhalo
disruption is supported by a larger number of dark matter-only simulations of
Milky Way-like hosts, run with (and without) an evolving disk potential \citep{kelley18}.

To mimic the disruption of subhalos in ELVIS, we adopt
$(N_{\rm DMO}/N_{\rm HYDRO})^{-1}$ as the likelihood that a subhalo survives to
$z=0$ as a function of pericentric distance; for
$d_{\rm peri} \ge 50~{\rm kpc}$, we assume no subhalo destruction
(i.e.~$N_{\rm HYDRO}/N_{\rm DMO} = 1$).
Within the ELVIS halo catalogs, we then randomly destroy subhalos as a function
of their pericentric distance given this probability of survival. 
In total, this removes approximately $25\%$ of the subhalo population at the
selected mass scale ($M_{\rm peak} = 10^{7.9-9.75}~\msun$).
Throughout the remainder of this work, we refer to these modified halo
populations as comprising the ``Fat'' ELVIS halo catalogs, given their inclusion
of the destructive effects produced by the host's additional baryonic mass
component.
As hosts of the Milky Way's UFD population, we select subhalos from our Fat
ELVIS catalogs at $z = 0$ within the host virial radius and within a mass range
of $M_{\rm peak} = 10^{7.9-9.75}~\msun$, following the stellar mass-halo mass
(SMHM) relation of \citet{gk14}. 
This yields a population of $15,269$ subhalos across the $48$ ELVIS host systems.

The ELVIS merger trees include $75$ snapshots ranging from $z = 125$ to $z =
0$. Following \citet{fham15}, all halo properties are spline interpolated across
the snapshots at a time resolution of $20$~Myr, which enables more precise
measurement of subhalo infall times and pericentric distances.
To constrain the infall time ($t_{\rm infall}$) for each subhalo in our Fat
ELVIS catalogs, we measure the redshift at which a subhalo was first and last
accreted onto its host halo. In $51\%$ of cases, the first infall is the only
infall, such that $t_{\rm first} = t_{\rm last}$. To account for the potential
role of pre-processing, we also track the first infall onto any host halo with
$M_{\rm peak} \ge 10^{10.8}~\msun$ at $z = 0$. Following the SMHM relation of
\citet{gk14}, this host selection corresponds to systems that are similar to the
Small Magellanic Cloud (SMC) or more massive. In total, roughly $65\%$ of
subhalos in our chosen mass range ($M_{\rm peak} = 10^{7.9-9.75}~\msun$)
experience pre-processing, such that they are influenced by environment roughly
$2.4$~Gyr earlier on average \citep[see also][]{wetzel15a}.
Throughout this work, we take the last infall onto the current host (i.e.~onto a
Milky Way-like host) as the infall time for a subhalo, unless otherwise
stated. In general, our primary results are qualitatively independent of the
adopted definition of infall time.

As shown in Figure~\ref{fig:cumlaccrt}, the distribution of subhalo infall times
is very weakly dependent upon subhalo mass at $M_{\rm peak} < 10^{10}~\msun$,
such that our results are largely independent of the assumed stellar mass-halo
mass relation.
Likewise, we find very little difference in the distribution of infall
  times for subhalos associated with the Local Group-like, paired hosts versus
  the isolated hosts in the ELVIS suite, with subhalos typically accreted
  $<0.5$~Gyr earlier in the Local Group-like simulations.
In contrast, the typical infall time of a subhalo depends much more
significantly on host-centric distance, with those systems located near the host
biased towards early accretion.
For our sample of low-mass halos, the inclusion of tidal effects shifts the
distribution of subhalo infall times by $\sim 0.7$~Gyr earlier on average (see
black dash-dotted line in Figure~\ref{fig:cumlaccrt}).
Our fiducial subhalo population, selected to have
  $0.15 < R/R_{\rm vir} < 0.5$ and $10^{8.4} < M_{\rm peak}/\msun < 10^{9.2}$,
  includes a total of $1,739$ subhalos and is well-matched to the UFD sample of
  \citet{brown14} based on host-centric distance and stellar mass, as shown
  relative to the greater MW UFD population in Figure~\ref{fig:ufdsamp}.

\subsection{Methods}

We employ a simple statistical method to quantify the probability that
environmental mechanisms may be responsible for quenching star
formation in a given population of subhalos (i.e.~UFDs).
From the parent subhalo population, chosen to match a particular observed galaxy
sample, we select (with replacement) a sample of $N$ random subhalos. If all $N$
subhalos are accreted onto their host halo (for the last time) at or before a
given redshift, then for that redshift the entire set of subhalos is considered
quenched. This process is replicated across 10,000 trials at each $z$, spanning
from $z = 4$ to $z =0$ at intervals of $\Delta z = 0.05$. The ``environmental
quenching probability'' as a function of cosmic time (or $z$) is then calculated
as the ratio of trails where all $N$ systems quench relative to the total number
of trials (i.e.~10,000).
Throughout the remainder of this work, we explore the dependence of this
environmental quenching probability on the sample size ($N = 6, 10, 20$), the
adopted infall time (e.g.~allowing for pre-processing by lower-mass hosts), and
the fraction of the sample required to be quenched at a given redshift.

\begin{figure}
\centering 
\includegraphics[width=0.45\textwidth]{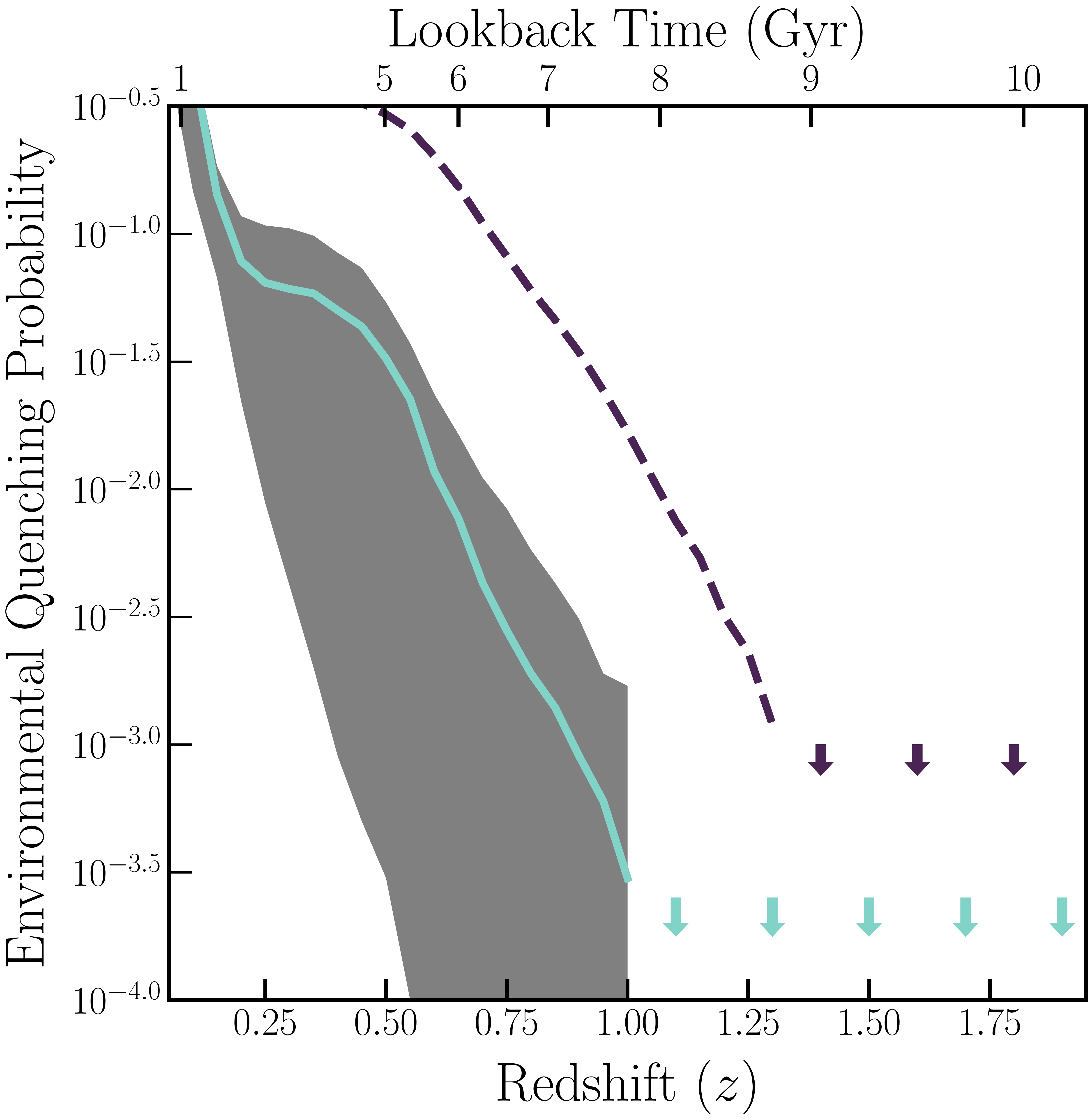}
\caption{The probability that a random sample of $6$ subhalos, selected as
  likely UFD hosts, were all accreted prior to a given redshift ($z$). The aqua
  line illustrates this ``environmental quenching probability'' as a function of
  redshift for our fiducial subhalo sample, while the grey shaded region
  illustrates the scatter associated with varying our selection of subhalos
  across the range $0.01 < R/R_{\rm vir} < 0.9$. The dashed plum line includes the
  role of pre-processing (infall onto a $\geq$ SMC-like host halo). The
  likelihood that environmental processes quenched the $6$ UFDs from
  \citet{brown14} is relatively small ($<1\%$). }
\label{fig:results}
\end{figure}

\section{Results}
\label{sec:results}

To determine if environmental effects were responsible for quenching the
present-day lowest-mass satellites of the Milky Way, we utilize our fiducial Fat ELVIS
subhalo population to constrain the likelihood that all $6$ galaxies in the
\citet{brown14} UFD sample were accreted at early cosmic times --- such that
environmental quenching could reproduce the observed SFHs of these systems.
From our fiducial subhalo sample, we randomly draw (with replacement) $6$
subhalos and evaluate -- as a function of redshift -- whether the entire sample
of $6$ was accreted by a given $z$.
Repeating this exercise across 10,000 trials, we compute the likelihood that a
sample of $6$ randomly-chosen UFDs could be environmentally quenched as a
function of cosmic time.

As shown in Figure~\ref{fig:results}, there is a vanishingly small probability
that $6$ random subhalos would all be accreted at high redshift (i.e.~$z > 1$)
or that the corresponding galaxies would be quenched by environmental process at
such early cosmic time.
At $z \sim 1$, after observations suggest that star formation halted in
  the UFD sample from \citet{brown14} including uncertainties of $\gtrsim1$~Gyr
  in the inferred SFHs, there is still an extremely low probability ($< 0.1\%$)
  that all $6$ systems could be quenched via environmental effects.
Allowing $\sim 1~{\rm Gyr}$ for a satellite to quench following infall
\citep{fham15}, such that all $6$ UFDs must be accreted by $z \sim 1.3$ to quench
by $z \sim 1$, only further decreases the potential impact of environmental
quenching (see Fig.~\ref{fig:results}).
While allowing for pre-processing in hosts down to SMC-like scales increases the
possible effectiveness of environmental effects (see dashed plum line in
Fig.~\ref{fig:results}), the likelihood that environment quenched the UFDs in
the \citet{brown14} sample is remarkably low ($<1\%$ for $z_{\rm quench} > 2$).
Overall, environmental mechanisms are unlikely to be responsible for the
universally old stellar populations inferred for the \citet{brown14} UFD sample.

\begin{figure}
\centering
\includegraphics[width=0.45\textwidth]{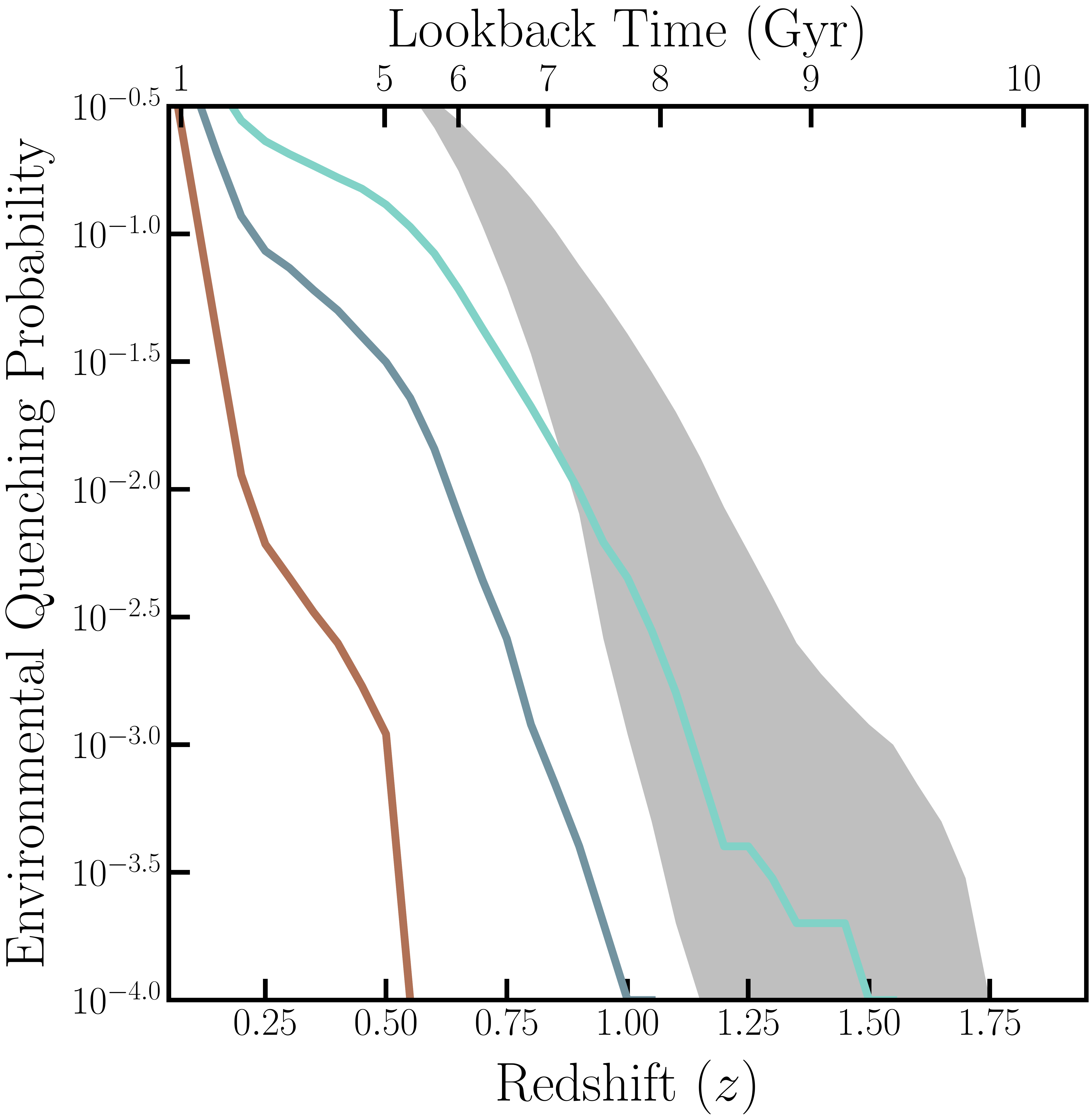}
\caption{The probability that all (\emph{solid lines}) or $75\%$ (\emph{shaded
    region}) of a random sample of $N$ subhalos, selected as likely UFD hosts,
  were accreted prior to a given redshift. For a parent subhalo population with
  $0.01 < R/R_{\rm vir} < 0.9$ and $10^{7.9} < M_{\rm peak}/\msun < 10^{9.75}$,
  the aqua, sage and sienna lines illustrate the environmental quenching
  probability as a function of redshift for subsamples of $N = 6, 10, 20$,
  representing our fiducial sample, the set of UFDs with SFHs, and the set of
  all UFDs with an estimated age, respectively.  The grey shaded region
  illustrates the environmental quenching probability for samples of $N=6$ to
  $N=20$ UFDs, requiring that only $75\%$ of the population was accreted by the
  given redshift.}
\label{fig:bigsamp}
\end{figure}

Including both \citet{brown14} and \citet{weisz14a}, there are published SFHs
for $10$ UFDs, all indicating that star formation halted by $z \gtrsim 2$.
Moreover, spectroscopic and/or photometric observations of (at least) a further
$10$ systems point to old (or metal-poor) stellar populations
\citep[e.g.][]{des2, laevens15, simon15, simon17, li18, torrealba18}. While
these additional UFDs span a broader range of galactocentric distance, with some
potentially pre-processed by the Magellanic Clouds \citep{koposov15, des1, des2,
  yonzin15, jethwa16, sales17}, the total sample of $20$ UFDs creates a powerful
dataset with which to examine the role of environment.
As expected, if we expand the sample of UFDs to all of those with well-measured
star-formation histories ($N = 10$) or yet larger to $N = 20$, it is even more
difficult to explain the universally-ancient stellar populations observed in
terms of an environmental effect.
Figure~\ref{fig:bigsamp} shows the probability that a sample of $N=10$ (sage
line) or $N=20$ (sienna line) UFD satellites were quenched following infall onto
the Milky Way halo as a function of cosmic time. We find that there is a
$\lesssim 0.01\%$ probability that samples of this size were entirely accreted
by $z = 2$. Even if we allow for late-time star formation in $25\%$ of the UFD
population (see grey shaded region in Fig.~\ref{fig:bigsamp}), we find that
environmental processes are unlikely to have been the dominant quenching
mechanism for the current sample of known UFDs orbiting the Milky Way.

%

\begin{figure*}
\centering
\includegraphics[width=1.\textwidth]{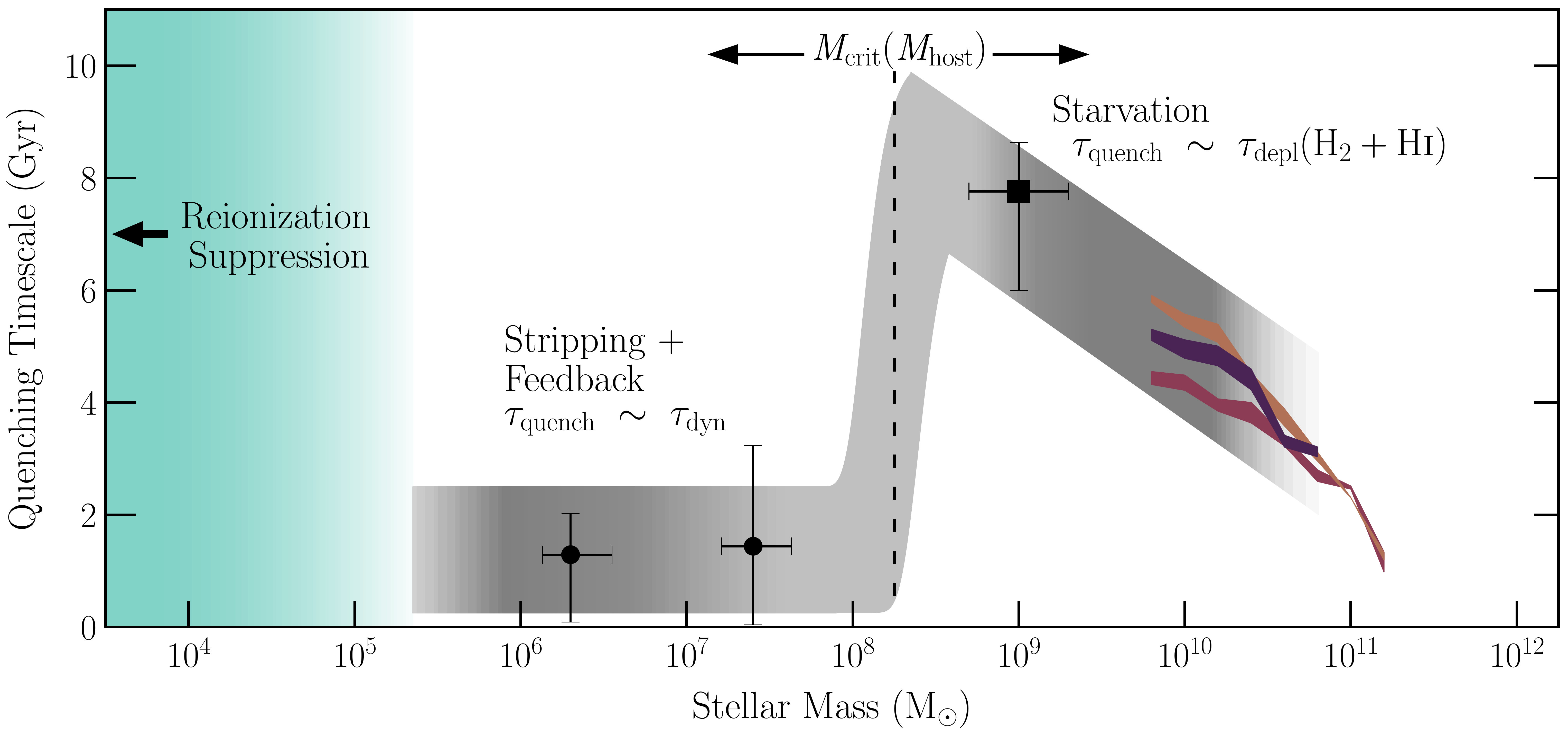}
\caption{The dependence of the satellite quenching timescale on satellite
  stellar mass in massive host halos ($\gtrsim10^{12}~\msun$), as adapted from
  \citet{fham15, fham16}. The plum, sienna, and burgundy colored bands show the
  constraints from \citet{wetzel13} for satellites in host halos of
  $M_{\rm host} \sim 10^{12-13}~\msun$, $10^{13-14}~\msun$, and
  $10^{14-15}~\msun$, respectively. The black square and circles correspond to
  the typical quenching timescale for intermediate- and low-mass satellites from
  \citet{wheeler14} and \citet{fham15}, respectively. The light grey shaded
  regions highlight the expected dominant quenching mechanism as a function of
  satellite mass, while the vertical dashed black line denotes the critical mass
  scale below which satellite quenching becomes increasingly efficient for a
  roughly Milky Way-like host. This critical mass, at which the dominant
  quenching mechanism changes, should increase with host halo mass. Finally, the
  aqua shaded region highlights the mass range where reionization is the most
  probable quenching mechanism.}
\label{fig:bigpic}
\end{figure*}

\section{Discussion}
\label{sec:dis}

\subsection{Quenching on the Smallest Scales}

Our analysis shows that the old stellar populations (and lack of significant
star formation at $z \lesssim 2$) observed in the Milky Way's UFD satellites is
unlikely to be reproduced via environmental quenching. Instead, the observed
star-formation histories of local UFDs are much more likely to have been
truncated via reionization.
Building upon the analysis of \citet{fham15, fham16}, Figure~\ref{fig:bigpic}
presents a complete picture of the dominant physical processes driving
late-time satellite quenching across more than $7$ orders of magnitude in
satellite stellar mass. In particular, we plot the current constraints on the
satellite quenching timescale (measured relative to infall) as a function of
satellite stellar mass; we caution that these measurements span a broad range of
host halo masses (from $\sim10^{12-15}~\msun$), but do describe a coherent
physical scenario (\citealt{wetzel13, wheeler14, fham15, fham16, fham18}, see
also \citealt{delucia12, hirschmann14, davies16}).

As illustrated in Fig.~\ref{fig:bigpic}, above a host-dependent critical mass
scale, satellites are able to largely resist stripping forces, such that they
are quenched on longer timescales consistent with starvation \citep{larson80,
  fham15}.
Below this critical mass scale, which is roughly $\mstar \sim 10^{8}~\msun$ for
Local Group-like hosts \citep{wheeler14, phillips15}, stripping is able to
remove the fuel supply for star formation from infalling satellites, such that
quenching occurs on roughly a dynamical time \citep{fham15, fham16, wetzel15b}.
This critical mass scale increases with host halo mass, such that stripping is
efficient at greater satellite masses in more massive host halos
\citep[e.g.][]{kenney89, solanes01, boselli14}; meanwhile, there likely exists
some limiting host mass (e.g.~$M_{\rm halo} \sim 10^{11}~\msun$) for which
stripping is inefficient on all mass scales and local environment is unable to
quench satellites ($\tau_{\rm quench} \sim \tau_{\rm depl} > t_{\rm hubble}$).
Finally, at the very lowest masses ($\mstar \lesssim 10^{5}~\msun$),
reionization acts to suppress star formation, independent of environment
(i.e.~for both isolated and satellite systems). We illustrate this regime in
Fig.~\ref{fig:bigpic} as the aqua shaded region.

Our results are consistent with recent hydrodynamical simulations of
  galaxy formation, which find that suppression of star formation by
  reionization is commonplace below a mass scale of
  $\mstar \lesssim 10^{5}~\msun$ \citep{bl15, fitts17, jeon17, dawoodbhoy18,
    aubert18}.
While reionization halts the infall of new gas in low-mass halos, residual star
formation can be fueled by the galaxy's existing gas reservoir so as to produce
star-formation histories similar to those observed for UFDs \citep{onorbe15,
  wheeler15}.
Additionally, reignition of star formation after initial suppression via
reionization may produce short and late periods of star formation
\citep{ledinauskas18, wright18}, such as that observed in Carina by
\citet{weisz14a}.
Observations in the Local Volume also broadly suggest that the mass scale at
which quenching via reionization dominates is approximately
$\mstar \sim 10^{5}~\msun$ \citep[e.g.][]{tp17}.
In particular, Leo~T has a stellar mass of $\mstar \sim 10^{5.5}~\msun$, with a
significant neutral gas reservoir \citep{rw08, adams17} and a complex
star-formation history, including significant activity at $z < 1$
\citep{dejong08, clementini12, weisz12}.
At a distance of $>400~{\rm kpc}$ from the Milky Way \citep{irwin07}, Leo~T
likely represents the tail of the star-forming field population, having a dark
matter halo mass greater than that at which reionization suppresses gas
cooling. Studies of stellar and gas kinematics in Leo~T suggest a halo mass of
$\sim10^{9}~\msun$ \citep{sg07, rw08}. 
And XVI \citep{ibata07}, a satellite of M31 with a stellar mass of
$\gtrsim10^{5}~\msun$ and a star formation history that extends to $z \sim 0.5$
\citep{weisz14c, monelli16}, was potentially a similar system prior to being
accreted by M31 and quenched via environmental mechanisms. 
While Leo~T and And XVI support a mass scale for quenching via reionization of
$\mstar \sim 10^{5}~\msun$ ($M_{\rm halo} \sim 10^{9}~\msun$), recent
observations of additional low-mass satellites of M31 indicate that the relevant
mass scale may be yet lower \citep[$\mstar \sim 10^{4.5}~\msun$,][]{martin16,
  martin17}.
It is important to note that there is likely not a clearly defined
stellar mass scale at which reionization is effective, given the potentially
large scatter in the stellar mass-halo mass relation at low masses
\citep{gk17b}. 

Taking $\mstar \sim 10^{5}~\msun$ ($M_{\rm halo} \sim 10^{9}~\msun$) as the
scale at which reionization suppresses star formation across all environments,
we predict a population of $\gtrsim 250$ UFDs within $1 < R/\rvir < 2$ of the
Milky Way and M31, based on counts of halos with
$M_{\rm halo} = 10^{7.9 - 9.75}~\msun$ in the Fat ELVIS catalogs across all 36
simulations.\footnote{On average, the $12$ paired host simulations have slightly
  more halos in the $1 < R/\rvir < 2$ range and a smaller fraction of these
  being backsplash halos.}  All of these systems are expected to be dominated by
ancient stellar populations. While some will have interacted with the Milky Way
and/or M31, a relatively large fraction ($>50\%$) of halos at these distances
are true ``field'' systems, having never spent time as a subhalo.
Future imaging surveys, such as the Large Synoptic Survey Telescope
\citep{ivezic08}, are expected to discover much of this population in the coming
decade, opening new avenues to study the suppression of star formation on the
smallest scales.
The total number of field UFDs will not only depend on the mass scale at which
reionization suppresses ongoing star formation at high $z$, but also the yet
lower scale at which it is able to suppress all star formation
\citep[e.g.][]{bullock00, somerville02}.

\subsection{The Curious Case of Eri~II}

If reionization truly quenches all low-mass galaxies, independent of
environment, we would expect that isolated UFDs should host ancient stellar
populations similar to those observed for known UFD satellites. The recent
discovery of Eridanus~II at a distance of $\gtrsim 350~{\rm kpc}$ from the Milky
Way \citep{des1, li17} has offered the opportunity to probe the SFH of a
``field'' UFD in significant detail. At a galactocentric distance of
$\sim 1.2~R_{\rm vir}$, however, Eri~II cannot be considered an isolated system,
unaffected by potential environmental effects. A significant fraction of systems
at such distances are associated with ``backsplash'' halos \citep{teyssier2012,
  gk14, fham18}, which previously passed within the host's (i.e.~Milky Way's)
virial radius before returning to the field.

\begin{figure}
\centering
\includegraphics[width=0.45\textwidth]{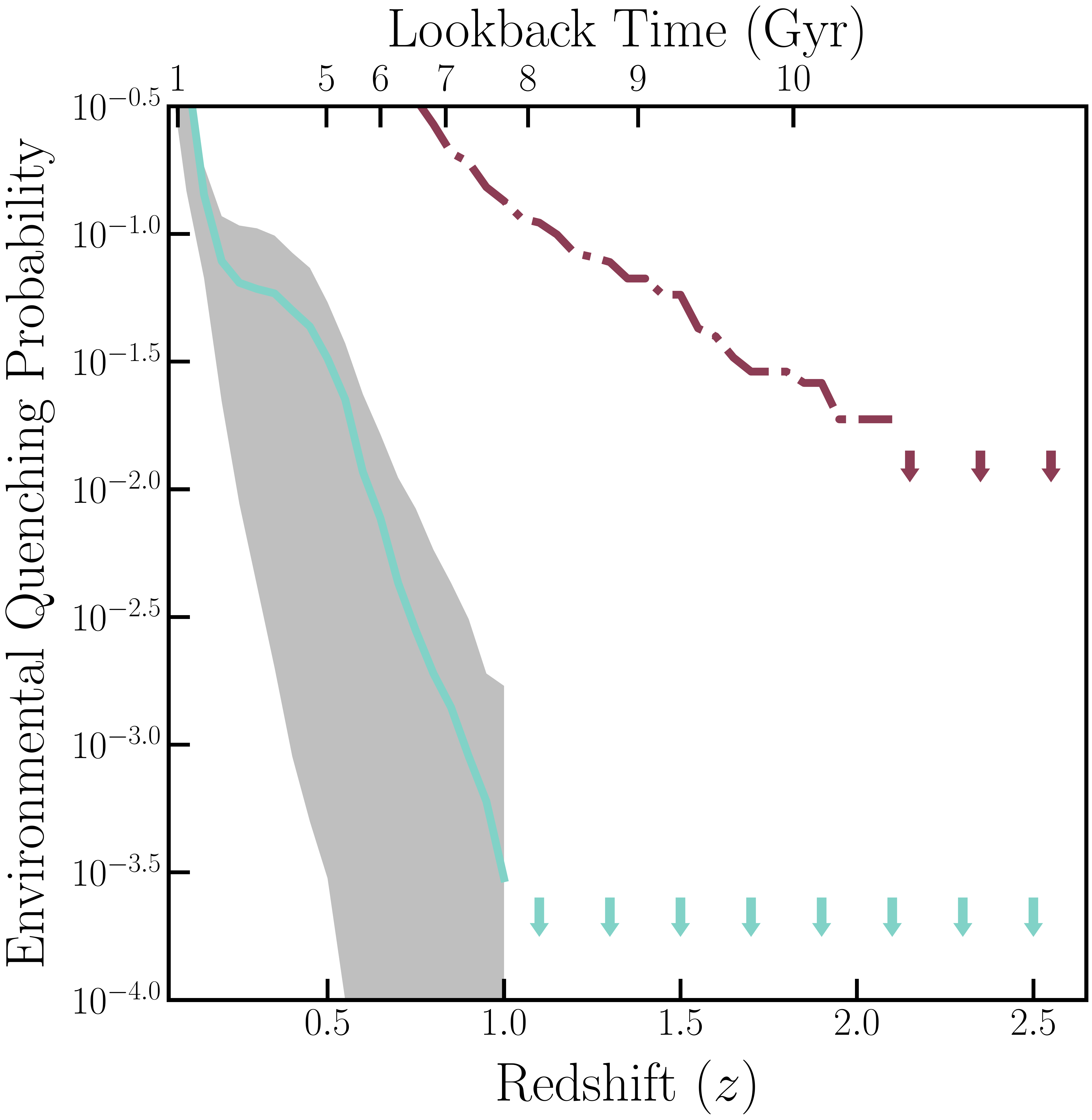}
\caption{The probability that a randomly-selected Eri~II-like halo was accreted
  by the Milky Way as a function of cosmic time (burgundy dash-dotted line). For
  comparison, we overplot the probability that a sample of $6$ subhalos were
  accreted by the same redshift (from Fig.~\ref{fig:results}). While Eri~II is
  unlikely to have been quenched by environment, the ancient stellar populations
  observed in current samples of UFD satellites argue more strongly against
  environment's role in suppressing star formation on the smallest scales.}
\label{fig:eri2}
\end{figure}

While recent observations show no signs of late-time star formation
(\citealt{li17}, but see also \citealt{koposov15, crnojevic16}), Eri~II -- as a
solitary system with an unknown orbital history -- places limited constraints on
the dominant mechanism responsible for suppressing star formation on the
smallest scales.
As shown in Figure~\ref{fig:eri2}, the current sample of Milky Way UFD
satellites already places a stronger constraint on the role of environment.
To test whether Eri~II is likely to have been quenched by environment, we select
subhalos from our Fat ELVIS catalogs, matching the mass
($8.9 < M_{\rm peak}/\msun < 9.75$), host-centric line-of-sight velocity
($-90~{\rm km}~{\rm s}^{-1} <V_{ \rm \ell os}< -40~{\rm km}~{\rm s}^{-1}$), and
host-centric distance ($0.9 < R/R_{\rm vir} < 1.9$) of Eri~II
\citep{li17}.\footnote{The adopted phase-space range was selected to encompass
  velocity and distance errors, as well as a possibly higher than originally
  assumed total velocity, following suit based on recently-derived velocities
  for UFDs from Gaia Data Release 2 \citep{simon18, fritz18}.}
From the resulting sample of $274$ subhalos, we compute the infall distribution
as a function of cosmic time (see Fig.~\ref{fig:eri2}), which corresponds to the
likelihood that environment played a role in quenching star formation in Eri~II.
We find that there is a $\sim10\%$ chance that Eri II was quenched via an
interaction with the Milky Way at $z \sim 1$.
While Eri~II is unlikely to have been quenched due to an interaction with the
Milky Way at $z > 2$ (so as to produce a purely old stellar population), the
measured SFHs for the existing sample of UFD satellites orbiting the Milky Way
already argue more strongly against environment's role in suppressing star
formation on the smallest scales.

\section{Summary}
\label{sec:sum}

Using the ELVIS suite of Milky Way- and Local Group-like $N$-body simulations to
constrain the infall times for subhalos likely to host the ultra-faint satellite
population of the Milky Way, we explore the potential role of environment in
suppressing star formation on small scales. Our principal results are as
follows.

\begin{itemize}[leftmargin=0.2in]

\item When incorporating the effects of subhalo tidal disruption due to the
  inclusion of the host's baryonic component, we find a shift in the typical
  infall time of $\sim 0.7$~Gyr for subhalos in the mass range of
  $M_{\rm halo} = 10^{7.9-9.75}~\msun$, such that subhalos are preferentially
  accreted at later cosmic time versus the same subhalos in a pure dark
  matter-only, $N$-body simulation. \\

\item For the $6$ UFDs included in the \citet{brown14} sample, we find that
  there is a $\lesssim 0.1\%$ probability that the Milky Way environment was
  solely responsible for quenching their star formation at $z > 1$. \\

\item For larger samples of UFDs, the likelihood that environment plays a
  dominant role in quenching decreases dramatically, such that there is a
  $< 0.01\%$ probability that environmental mechanisms are responsible for
  quenching all $10$ UFDs included in the \citet{brown14} and \citet{weisz14a}
  samples.\\

\item Given the inability of environmental effects to reproduce the observed
  star-formation histories of observed UFDs, we conclude that reionization is
  the most likely mechanism by which star formation is suppressed on the
  smallest scales. \\

\item Finally, we predict that there is a population of $\gtrsim 250$ UFDs
  within $1< R/\rvir < 2$ of the Milky Way and M31, all with ancient stellar
  populations. Future imaging surveys, such as LSST, will be able to uncover much
  of this population. \\

\end{itemize}

Combined with results from \citet{fham15} and \citet{fham16}, our results
produce a coherent physical picture describing the dominant quenching mechanism
across the entire range of satellite (and host) masses (see
Fig.~\ref{fig:bigpic}). At the very smallest scales, we argue that the
suppression of star formation is largely independent of environment and set by
the minimum halo mass at which reionization curtails gas accretion.

\section*{acknowledgements}
We thank Tyler Kelley, Dan Weisz, Josh Simon, Alex Riley, and Mary Jenkins for
helpful discussions regarding this project.
This work was supported in part by NSF grants AST-1518257, AST-1518291,
AST-1517226, and AST-1815475. MBK also acknowledges support from NSF CAREER
grant AST-1752913 and NASA grant NNX17AG29G. Additional support for this work
was provided by NASA through grants GO-12914, AR-13888, AR-13896, GO-14191,
AR-14282, AR-14289, AR-14554, and AR-15006 from the Space Telescope Science
Institute, which is operated by the Association of Universities for Research in
Astronomy, Inc., under NASA contract NAS 5-26555.
MKRW acknowledges support from the National Science Foundation Graduate Research
Fellowship. This material is based upon work supported by the National Science
Foundation Graduate Research Fellowship Program under Grant No.~DGE-1321846.
This research made use of {\texttt{Astropy}}, a community-developed core Python
package for Astronomy \citep{astropy13}. Additionally, the Python packages
{\texttt{NumPy}} \citep{numpy}, {\texttt{iPython}} \citep{ipython},
{\texttt{SciPy}} \citep{scipy}, and {\texttt{matplotlib}} \citep{matplotlib}
were utilized for the majority of our data analysis and presentation.

\bibliography{ufd}

\label{lastpage}

\end{document}